# The bat coronavirus RmYN02 is characterized by a 6-nucleotide deletion at the S1/S2 junction, and its claimed PAA insertion is highly doubtful

Yuri Deigin[1] and Rossana Segreto[2]

[1]Youthereum Genetics Inc., Toronto, ON Canada.

ydeigin@gmail.com

[2]Department of Microbiology, University of Innsbruck, Austria.

Rossana.Segreto@uibk.ac.at

## Abstract

Zhou et al. reported the discovery of RmYN02, a strain closely related to SARS-CoV-2, which is claimed to contain a natural PAA amino acid insertion at the S1/S2 junction of the spike protein at the same position of the PRRA insertion that has created a polybasic furin cleavage site in SARS-CoV-2. The authors support with their findings the theory that the furin cleavage site insertion present in SARS-CoV-2 is natural. Because no nucleotide alignment with closely related strains of the region coding for the supposed insertion is provided by Zhou et al., we have applied several alignment algorithms to search for the most parsimonious alignments. We conclude that RmYN02 does not contain an insertion at the S1/S2 junction when compared to its closest relatives at the nucleotide level, but rather a 6-nucleotide deletion and that the claimed PAA insertion is more likely to be the result of mutations. A close examination of RmYN02 sequencing records and assembly methods is wishful. In conclusion, SARS-CoV-2, with its 12-nucleotide insertion at the S1/S2 junction remains unique among its sarbecovirus relatives.

Recently, Zhou et al.[1] reported the discovery of a novel coronavirus strain RmYN02, which the authors claim to contain a natural PAA amino acid insertion at the S1/S2 junction of the spike protein at the same position of the PRRA insertion that has created a polybasic furin cleavage site in severe acute respiratory syndrome coronavirus type 2 (SARS-CoV-2).

Zhou et al. have come to their conclusion based on a multiple sequence alignment of RmYN02 with several beta coronavirus strains, namely SARS-CoV-2, SARS-CoV GZ02, RaTG13, ZC45, ZXC21, Pangolin/GD/2019, and Pangolin/GX/P5L/2017. Their findings are reported in an amino acid alignment diagram where the supposed PAA amino acid insertion is placed between the 680 (serine) and 685 (arginine) amino acids of SARS-CoV-2's spike protein.

However, no nucleotide alignment of the same region is provided by Zhou et al. that would allow the reader to identify the underlying nucleotides (CCT GCA GCG) coding the claimed PAA insertion in RmYN02 in relation to the other strains analyzed. We have thus performed a CLUSTAL W multiple nucleotide sequence alignment of the strains reported in Zhou et al., but were unable to observe the claimed insertion (Fig. 1A). RmYN02 seems instead to contain a 6-nucleotide deletion at the S1/S2 junction when compared to the other strains, and the only insertion observed when aligning the genomes in question is the well-known 12-nucleotide insertion CT CCT CGG CGG G (PRRA) into SARS-

CoV-2. The 6-nucleotide deletion in RmYN02 at the S1/S2 junction is even more apparent when SARS-CoV-2 is excluded from the multiple sequence alignment (Fig. 1B).

We believe that including SARS-CoV-2 in the alignment together with RmYN02 and other strains is methodologically incorrect, as the implied underlying hypothesis which the analysis is meant to test is that SARS-CoV-2's PRRA insertion is of natural origin. Thus, including SARS-CoV-2 in the alignment not only biases the alignment algorithm, but also pre-supposes the conclusion that the PRRA insert is, indeed, natural. To prove that inserts like PRRA occur naturally, strains that exhibit similar inserts must be compared to their relative strains, excluding SARS-CoV-2 from the analysis. Our analyses show that RmYN02 does not contain an insertion at the S1/S2 junction when compared to its closest relatives and the claimed PAA insertion is more likely to be the result of mutations. Pairwise comparison between RmYN02 and its closest relatives (RaTG13, ZC45, ZXC21) confirms this conclusion, when either RmYN02 (Fig. 1C) or ZC45 (Fig. 1D) is used as an anchor, and instead produces a 2-nt deletion in the coding region for PAA (Fig. 1D). If RmYN02 truly had an insertion comparable to the PRRA insertion in SARS-CoV-2, we would have expected such an insertion to be clearly observable in pairwise comparisons to its closest relatives, which are RaTG13, ZC45, ZXC21, and Pangolin/GD/2019 (Fig. 1E).

A close examination of the S1/S2 region reveals that in RmYN02 it is 6 nucleotides (2 amino acids) shorter than those of its related strains RaTG13, Pangolin/GD/2019, ZC45, and ZXC21. Therefore, to support the claimed PAA insertion not only a 9-nucleotide insertion, but also a 15-nucleotide deletion must have occurred. While this is theoretically possible, we propose 3 alternatives of more parsimonious alignments which either do not have any insertions (ver. 1 and ver. 3 in Fig. 1F), or at most have a 3-nucleotide insertion (ver. 2 in Fig. 1F). The alignment proposed by CLUSTAL W also did not produce any insertions (ver. "Clustal W" in Fig. 1F).

Rather than a complete 12-nucleotide deletion of the region in RmYN02 that corresponds to QTQT in RaTG13 as proposed by Zhou et al., a more parsimonious scenario is a 3-nucleotide deletion split between the first and fourth codons of QTQT, thereby turning it into NSP in RmYN02. Another possibility, proposed by CLUSTAL W, is a 6-nucleotide deletion in the middle of the nucleotides coding for QTN, turning it into a P.

On the other side of the PAA(R) insertion claimed by Zhou et al., we feel that a more parsimonious alignment of RmYN02 is best elucidated via comparing it to its close relative strains ZC45 and RaTG13: in particular, the **CGC** A**GT** nucleotides in ZC45 coding for RS align best to the G**CG CGT** nucleotides in RmYN02, having possibly resulted from an insertion of G and deletion of A nucleotides in RmYN02 relative to ZC45 (Fig. 1G).

Finally, we should not discount the possibility that the observed 6-nucleotide deletion in RmYN02 at the S1/S2 junction, so uncharacteristic of its relative strains, is a result of sequencing error rather than a bona fide deletion.

While further virus collecting expeditions might produce unanticipated discoveries, to date SARS-CoV-2 remains unique among its sarbecovirus relatives not only due to a polybasic furin site at the S1/S2 junction, but also due to the length of the locus surrounding the 12-nucleotide insert that has created the furin site: SARS-CoV-2 is at least 12 nucleotides longer at that junction than any of its sarbecovirus relatives. Its PRRA insertion is beyond any doubts, and was not accompanied by any deletions, which stands in sharp contrast to what is observed in RmYN02.

In closing, we would like to mention that the RmYN02 sequence is presently only available in the GISAID database, which is password protected and requires registration. We would propose that RmYN02 should also be made available at GenBank.

# Conflicts of Interest Statement

RS and YD do not have any conflicts of interest.

# Figure

*Figure 1:*

- (A) Multiple sequence alignment of RmYN02 with the strains used in Zhou et al. for comparison. RmYN02's nucleotides coding the PAA amino acids (CCT GCA GCG) are surrounded by a red box. No insertion in RmYN02 is visible; on the contrary, a deletion splitting the nucleotides coding for PAA is observed.

- (B) Multiple sequence alignment of RmYN02 with the strains used in Zhou et al. for comparison, with the exception of SARS-CoV-2. The deletion characterizing RmYN02 at the S1/S2 junction appears to cause a split of the first nucleotide from the rest of the sequence coding for the PAA amino acids (CCT GCA GCG, surrounded by a red box).

- (C) Pairwise comparisons of RmYN02 (anchor) to RaTG13, ZC45, and ZXC21. No PAA insertion is observed in RmYN02 in these comparisons.

- (D) Pairwise comparisons of ZC45 (anchor) to ZXC21, RmYN02, RaTG13, and Pangolin/GD/2019. RmYN02's nucleotides coding the PAA amino acids (surrounded by a red box) are aligned as mutations relative to ZC45 rather than insertions.

- (E) Phylogenetic tree of SARS-GZ02, Pangolin/GX/2017, ZC45, ZXC21, RmYN02, RaTG13, and Pangolin/GD/2019 produced by CLUSTAL W based on the alignment of their genomes as in Fig. 1B.

- (F) Nucleotide and amino acid alignments of RmYN02 with SARS-CoV-2, RaTG13, Pangolin/GD/2019, RmYN01, RP3, Rf4092, LYRa11, Rs3367, RsSHC014, ZC45, and ZXC21 at the S1/S2 junction of the spike protein. For RmYN02, three alternative versions are provided, besides the ones proposed by Clustal W and Zhou et al.

- (G) Nucleotide and amino acid alignments of RmYN02 with its closely related strains RaTG13 and ZC45 that better illustrates potential mutations and deletions between these strains, but no PAA insertion.

**1A**

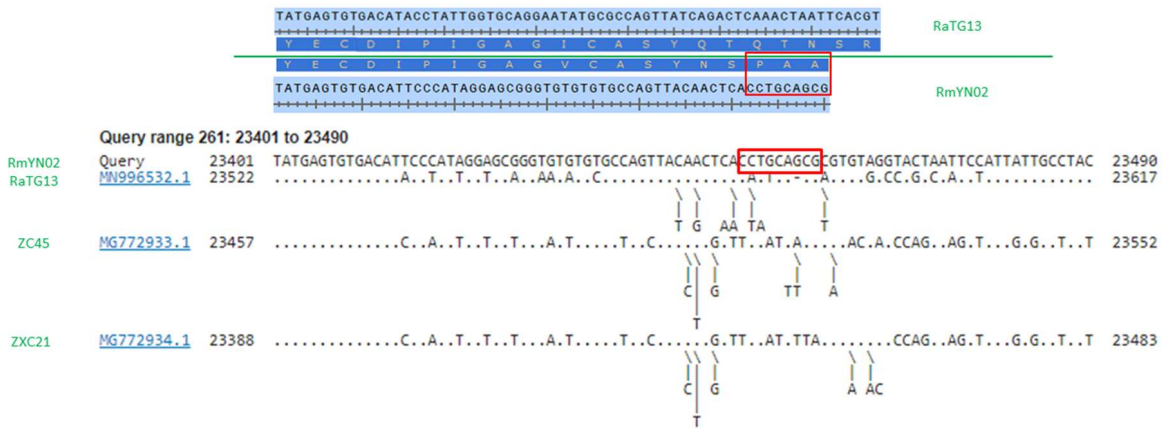

**1B**

**1C**

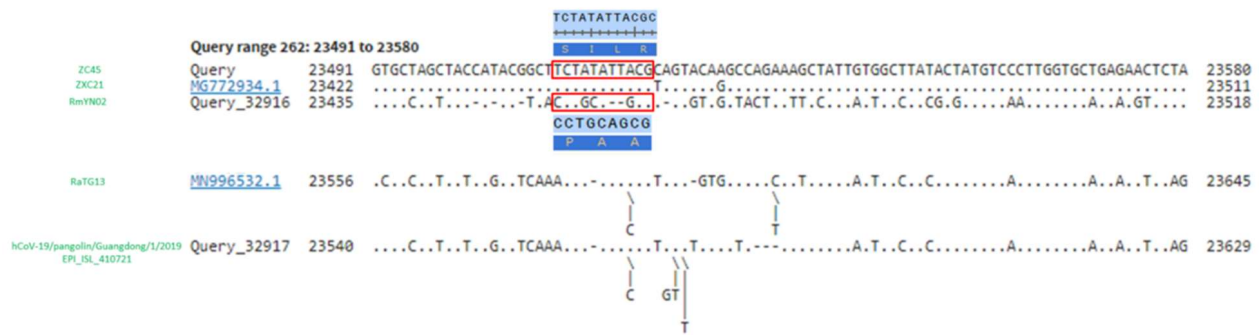

**1D**

## 1E

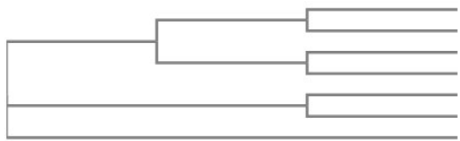

```
              ┌─ GZ02_AY390556.1 0.12708
              ├─ pangolin/Guangxi/P2V/2017|EPI_ISL_410542 0.08268
              ├─ ZC45_MG772933.1 0.01304
              ├─ ZXC21_MG772934.1 0.01225
              ├─ RmYN02 0.03996
              ├─ RaTG13_MN996532.1 0.02907
              └─ hCoV-19/pangolin/Guangdong/1/2019|EPI_ISL_410721 0.054
```

## 1F

| | | |
|---|---|---|
| SARS-CoV-2 | G A G I C A S Y Q T Q T N S P R R A R S V A S Q S I I | ggt gca ggt ata tgc gct agt tat cag act cag act aat tct cct cgg cgg gca cgt agt gta gct agt caa tcc atc att |
| RaTG13 | G A G I C A S Y Q T Q T N S - - - - R S V A S Q S I I | ggt gca gga ata tgc gcc agt tat cag act caa act aat tca --- --- --- --- cgt agt gtg gcc agt caa tct att att |
| Pangolin/GD/2019 | G A G I C A S Y Q T Q T N S - - - - R S V S S Q A I I | ggt gca gga ata tgt gcc agt tat cag act caa act aat tca --- --- --- --- cgt agt gtt tca agt caa gct att att |
| Pangolin/GX/2017 | G A G I C A S Y H S M S S F - - - - R S V N Q R S I I | ggt gct ggc ata tgt gca agt tac cat tcc atg tca ttt --- --- --- --- cgt agt gtc aac cag tca atc att |
| RmYN02 Zhou et al. | G A G V C A S Y - - - - N S P - A A R - V G T N S I I | gga gcg ggt gtg tgt gcc agt tac --- --- --- --- aac tca cct --- gca gcg cgt --- gta ggt act aat tcc att att |
| RmYN02 Clustal W | G A G V C A S Y N S ( P ) A - - - - A R V G T N S I I | gga gcg ggt gtg tgt gcc agt tac aac tca c-- --- -ct gca --- --- --- --- gcg cgt gta ggt act aat tcc att att |
| RmYN02 ver 1 | G A G V C A S Y (N S P) - A - - - - A R V G T N S I I | gga gcg ggt gtg tgt gcc agt tac -a- act cac -ct --- gca --- --- --- --- gcg cgt gta ggt act aat tcc att att |
| RmYN02 ver 2 | G A G V C A S Y (N S P) - A - - - A R - V G T N S I I | gga gcg ggt gtg tgt gcc agt tac -a- act cac -ct --- gca --- --- --- --- gcg cgt --- gta ggt act aat tcc att att |
| RmYN02 ver 3 | G A G V C A S Y N S P A - - - - - A R V G T N S I I | gga gcg ggt gtg tgt gcc agt tac aac tca cct gca --- --- --- --- --- gcg cgt gta ggt act aat tcc att att |
| RmYN01 | G A G I C A S Y H T A S L L - - - - R N T G Q K S I V | ggt gca ggc att tgt gct agt tac cat aca gct tcc ctt tta --- --- --- --- cgt aat aca ggc cag aaa tcc att gtg |
| RP3 | G A G I C A S Y H T A S T L - - - - R S V G Q K S I V | ggt gct ggc att tgt gct agc tac cat aca gct tct act tta --- --- --- --- cgt agt gta ggt cag aaa tcc att gtg |
| Rf4092 | G A G I C A S Y H T A S T L - - - - R G V G Q K S I V | ggt gct ggc att tgt gct agc tac cat aca gct tct act cta --- --- --- --- cgt ggt gta ggt cag aaa tcc att gtg |
| LYRa11 | G A G I C A S Y H T A S L L - - - - R N T D Q K S I V | ggt gct ggc att tgt gct agt tac cat aca gct tct ctc tta --- --- --- --- cgt aat aca gac aaa tca att gtg |
| Rs3367 & RsSHC014 (identical here) | G A G I C A S Y H T V S S L - - - - R S T S Q K S I V | gga gct ggc att tgt gct agt tac cat aca gtt tct tca tta --- --- --- --- cgt agt act agc caa aaa tct att gtg |
| ZC45 | G A G I C A S Y H T A S I L - - - - R S T S Q K A I V | ggt gct ggt att tgt gct agc tac cat acg gct tct ata tta --- --- --- --- cgc agt aca agc cag aaa gct att gtg |
| ZXC21 | G A G I C A S Y H T A S I L - - - - R S T G Q K A I V | ggt gct ggt att tgt gct agc tac cat acg gct tct ata tta --- --- --- --- cgt agt aca ggc cag aaa gct att gtg |

```
Black   = common for all
Purple  = unique to SARS-CoV-2
Blue    = differences mostly found in ZC45 or ZXC21
Green   = differences mostly found in RmYN02 or RmYN01
Yellow  = differences mostly found in Pangolin/GX/2017
Red     = other differences
```

## 1G

| | | |
|---|---|---|
| RaTG13 | G A G I C A S Y Q T Q T N S R S V A S Q S I I | ggt gca gga ata tgc gcc agt tat cag act caa act aat tca cgt agt gtg gcc agt caa tct att att |
| RmYN02 ver 1 | G A G V C A S Y (N S P) - A A R V G T N S I I | gga gcg ggt gtg tgt gcc agt tac -a- act cac -ct --- gca (gcgc -gt)gta ggt act aat tcc att att |
| ZC45 | G A G I C A S Y H T A S I L R S T S Q K A I V | ggt gct ggt att tgt gct agc tac cat acg gct tct ata tta cgc agt aca agc cag aaa gct att gtg |

```
Black  = common for all three strains
Green  = unique to RmYN02 among the three
Red    = unique to RaTG13 among the three
Blue   = unique to ZC45 among the three, or shared by ZC45 and RmYN02
```